\documentclass[conference]{IEEEtran}
\ifCLASSINFOpdf
\else
\fi
\IEEEoverridecommandlockouts
\usepackage{amssymb}
\usepackage{graphicx}
\usepackage{amssymb}
\usepackage{amsmath}
\usepackage{color,rotating}
\usepackage{lmodern}
\usepackage{url}

\newtheorem{Theorem}{Theorem}

\newcommand{\cqfd}{\mbox{}\nolinebreak\hfill\rule{2mm}{2mm}\medbreak\par}
\newcommand{\by}{\mbox{\boldmath $y$}}
\newcommand{\bv}{\mbox{\boldmath $v$}}

\newlength{\myfigwidth}
\begin{document}

\title{Optimal Scanning Bandwidth Strategy Incorporating
Uncertainty about Adversary's Characteristics\thanks{
This is the last draft version of the paper. Revised version of the paper was published in EAI Endorsed Transactions on Mobile Communications and Applications, Vol. 14, Issue 5, 2014, doi=10.4108/mca.2.5.e6}}

\author{\IEEEauthorblockN{Andrey
Garnaev} \IEEEauthorblockA{WINLAB, Rutgers University\\
North Brunswick, USA \\
Email: garnaev@yahoo.com} \and \IEEEauthorblockN{Wade Trappe}
\IEEEauthorblockA{
WINLAB, Rutgers University \\
North Brunswick, USA\\
E-mail: trappe@winlab.rutgers.edu}
}

\maketitle

\begin{abstract}
In this paper we investigate the problem of designing a spectrum scanning strategy to detect an
 intelligent Invader who wants to utilize spectrum
undetected for his/her unapproved purposes.  To deal with this
problem we  model the situation as
two games, between a Scanner and an Invader, and solve them sequentially. The first game is formulated  to  design the optimal (in maxmin sense) scanning algorithm, while the second one  allows one to find the optimal values of the parameters for the algorithm depending on parameters of the network.   These games provide solutions for two dilemmas that the rivals face.   The Invader's dilemma consists of the following: the more bandwidth the Invader attempts to use leads to a
larger payoff if he is not detected, but at the same time also
increases the probability of being detected and thus fined.
Similarly, the Scanner faces a dilemma: the wider the bandwidth
scanned, the higher the probability of detecting the Invader, but
at the expense of increasing the cost of building the scanning
system.  The equilibrium strategies are found explicitly and
reveal interesting properties. In particular, we have found a
discontinuous dependence of the equilibrium strategies on the
network parameters, fine and the type of the Invader's award. This
discontinuity of the fine means that the network provider has to take
into account a human/social factor since some threshold values of fine
could be very sensible for  the Invader, while in other situations
simply increasing the fine has minimal deterrence impact. Also we
show how incomplete information about the Invader's technical
characteristics and reward (e.g.  motivated by using
different type of application, say, video-streaming or downloading
files)   can be incorporated into scanning strategy to increase
its efficiency.
\end{abstract}


\section{Introduction}
Over the last few decades, the increasing demand for wireless
communications has motivated the exploration for more efficient
usage of spectral resources (\cite{1391031,819467}). In
particular, it has been noticed that there are large portions of
spectrum that are severely under-utilized \cite{akyildiz2006next}.
Recently, cognitive radio technologies (CR) have been proposed as
a means to intelligently use such spectrum opportunities by
sensing the radio environment and exploiting available spectrum
holes for secondary usage \cite{fette2009cognitive}. In CR systems,
secondary users are allowed to ``borrow (or lease)'' the usage of
spectrum from primary users (licensed users), as long as they do
not hinder in the proper operation of the primary users'
communications. Unfortunately, as we move to make the CR technologies
commercial, which will allow secondary users to access spectrum
owned by primary users, we will face the inevitable risk that
adversaries will be tempted to use CR technology for illicit and
selfish purposes \cite{liu2009aldo}. If we imagine an unauthorized
user (Invader) attempting to sneak usage of spectrum without
obeying proper regulations or leasing the usage of the spectrum,
the result will be that both legitimate secondary users and
primary users will face unexpected interference, resulting in
significant performance degradation across the system.

The challenge of enforcing the proper usage of spectrum requires
the notion of a ``spectrum policing agent'', whose primary job is
to ensure the proper usage of spectrum and identify anomalous
activities occurring within the spectrum\cite{liu2009aldo}.  As a
starting point to being able to police the usage of spectrum, we
must have the ability to scan spectrum and effectively identify
anomalous activities. Towards this objective, there have been
several research efforts in signal processing techniques that can
be applied to the spectrum scanning problem. For example, in
\cite{verdu1998multiuser,van2004detection}, the authors presented
methods for detecting a desired signal contained within
interference. Similarly, detection of unknown signals in noise
without prior knowledge of authorized users was studied in
\cite{digham2007energy,urkowitz1967energy}. As another example, in
\cite{liu2009aldo}, the authors proposed a method to detect
anomalous transmission by making use of radio propagation characteristics.
In \cite{GTK2012}, the authors investigated what
impact on spectrum scanning can have information about  the
over-arching application that a spectrum thief might try to run, while, in \cite{GT2013_a}, a stationary bandwidth scanning strategy in a discounted repeated game was suggested.

However, these pieces of work tend to not examine the important
``interplay'' between the two participants inherent in the
problem-- the Invader, who is smart and will attempt to use the
spectrum in a manner to minimize the chance of being detected and
fined, while also striving to maximize the benefit he/she receives
from illicit usage of this spectrum; and the Scanner, who must be
smart and employ a strategy that strategically maximizes the
chance of detecting and fining the smart Invader, with minimal
cost. This challenge is made more difficult by the complexity of
the underlying scanning problem itself: there will be large swaths
of bandwidth to scan, and the system costs (e.g., analog-to-digital
conversion, and the computation associated with running signal
classifiers) associated with scanning very wide bandwidth makes it
impossible to scan the full range of spectrum in a single
instance. Consequently, it is important to understand the
strategic dynamics that exist between the Scanner and the Invader,
while also taking into account the underlying costs and benefits
that exist for each participant as well as information or its lack
on the technical characteristics of the Invader and his object to
intrude into the bandwidth. This paper\footnote{The authors note
that a shortened version of this research was presented at
Crowncom 2013 \cite{GTK2013}, and this paper  extends the idea presented at Crowncom.} focuses on finding
the optimal scanning strategy by selecting the scanned (and,
similarly, the invaded) bandwidth that should be employed in
spectrum scanning and examining how incorporating information
or the lack of information about the technical characteristics of the Invader and
his object can improve the scanning strategy. In order to solve this
problem we will apply a Bayesian approach. Note that Bayesian
approaches have been widely employed in dealing with different
problems in networks, for example, intrusion  detection
\cite{LCM2006, ADBA2004, GGP2014}, scanning bandwidth \cite{GTK2012} and
transmission under incomplete information \cite{HMDH2009, AAG2011,
HDA2005,[13],[14],[15]}.
Finally note that the optimal scanning problem also relates the problem of
designing security systems. Note that an extensive literature exists on the construction and modeling of
different aspects of such security systems  for communication and
network security \cite{AB2003, NAB2009, AAA2009_gamenets, HMOS2005, RESDSW2010, GGG1997,
LCM2006, CMP2005, BG2000}, security in wireless networks \cite{MS2010,
ZSPB2011} and cyber-security \cite{KW2005, ADBA2004, HMDH2009}.  In
\cite{MZABH2013}, the readers can find a structured and comprehensive survey
of the research contributions that
analyze and solve security and privacy problems in computer networks via game-theoretic approaches.

The organization of this paper is as follows: in Section
\ref{sec:Problem}, we first define the problem by formulating
two  games, which will be solved sequentially  in terms of payoff and cost functions.  In the first
game, the Scanner looks for the maxmin scanning algorithm, if parameters (widths of used bandwidths)  of scanning and intrusion are fixed and known.
In the second
game each player, using the first game's result, which supplies detection probability,
looks for the optimal values of these parameters. To gain insight into the problem,
in Section \ref{sec:Linear}, we outline a linearized
model for detection probability and arrive at the corresponding
best response strategies for each player in Section \ref{sec:Best
Response_2}. We then explicitly obtain the equilibrium strategies,
 in Section \ref{sec:Nash Equilibrium_Unknown}   and Section \ref{sec:Nash
Equilibrium}, for cases involving complete and incomplete knowledge
of the Invader's technical characteristics (radio's capabilities). In Section
\ref{sec:Numerical illustrations}, numerical illustrations are
supplied. Finally, in Section \ref{sec:Discussion},  discussions and conclusions are supplied,  and,
in Appendix, the proofs of the
announced results  are offered to close the paper.

\section{\label{sec:Problem}
Formulation of the scanning problem}

In this section, we set up our problem formulation. Our formulation
of the spectrum scanning problem involves  two players: the
Scanner and the Invader. The Scanner, who is always present in the
system, scans a part of the band of frequencies that are to be
monitored, in order to prevent illegal usage by a potential
Invader of the primary (Scanner) network's ownership of this band.
We assume that the amount of bandwidth that needs to be scanned is
much larger than is possible using a single scan by the Scanner,
and hence the Scanner faces a dilemma: the more bandwidth that is
scanned, the higher the probability of detecting the Invader, but
at the expense of increasing the cost of the RF scanning system.

We assume that if the Scanner scans a particular frequency band
$I_S$ and the Invader uses the band $I_I$ then the invasion will
be detected with certainty if $I_S\cap I_I\not=\emptyset$, and it
will not be detected otherwise. Without loss of generality, we can
assume that the size of the protected frequency band is normalized
to 1. The Invader wants to use spectrum undetected  for some
illicit purpose. We consider two scenarios: (a) The reward for the
Invader is related to the width of the frequency band he uses if
he is undetected. If he is detected he will be fined. Thus, the
Invader faces a dilemma: the more bandwidth he tries to use yields
a larger payoff if he is not detected but also it increases the
probability of being detected and thus to be fined, (b) The reward
for the Invader is unknown to the Scanner: he only knows whether
it is related to the width of the frequency band the Invader uses,
or not. We formulate this problem as two games, which will be solved separately  in the
following two subsections.

\subsection{Formulation of the first
game - the scanning algorithm}

In the first game, where we look for a maxmin scanning algorithm,  the Scanner selects the band
$B_S=[t_S,t_S+x]\subseteq [0,1]$ with a fixed upper bound of
frequency width $x$ to scan i.e. $t_S\leq 1-x$. The Invader
selects the band $B_I=[t_I,t_I+y] \subseteq [0,1]$  with a fixed
upper bound frequency width $y$ to intrude, i.e., $t_I\leq 1-y$.
Thus, $B_S$ and $B_I$ are pure strategies for the Scanner and the
Invader. The Scanner's payoff $v(B_S,B_I)$ is 1 if the Invader is
detected (i.e. $[t_S,t_S+x]\cap [t_I,t_I+y]\not=\emptyset$) and
his payoff is zero otherwise. The goal of the Scanner is to
maximize his payoff, while the Invader wants to minimize it.
Thus, the Scanner and the Invader play a zero-sum game. The saddle
point (equilibrium) of the game is a couple of strategies
$(B_{S*},B_{I*})$ such that for each couple of strategies $(B_{S},B_{I})$
the following inequalities hold \cite{O1982}:
$$
v(B_{S},B_{I*})\leq v:=v(B_{S*},B_{I*})\leq v(B_{S*},B_{I}),
$$
where $v$ is the value of the game. It is clear that the game does
not have a saddle point in the pure strategy if $x+y<1$. To find
the saddle point we have to extend the game by mixed strategies,
where we assign a probability distribution over pure strategies.
Then instead of the payoff $v$ we have its expected value. The
game has a saddle point in mixed strategies, and let $P(x,y)$ be
the value of the game. Then $P(x,y)$ is the maximal detection
probability of the Invader under worst conditions.

\subsection{Formulation of the second
game - the optimal parameters of the scanning algorithm}

In the second  game the rivals knowing their
equilibrium strategies from the first game as well as  detection
probability $P(x,y)$, want  to find the equilibrium frequency
widths $x$ and $y$. We here consider three sub-scenarios: (a) the
Invader's type is known: namely, it is known how the reward for
the Invader is related to the width of the frequency band he uses
if he is undetected, (b) the technical characteristics of  the Invader
are known: namely, it is known which frequency band  is
available for him to use, (c) the Invader's type is unknown: instead,
there is only a chance that the Invader reward is related to the
width in use. Otherwise, it is not related. Different types of rewards
can be motivated by using different types of applications (say,
file-download or streaming video).

\subsubsection{Invader reward is related to the bandwidth used}
\label{sec:Invader reward is related to the bandwidth used}
 A  strategy for the Scanner is to scan a width of
frequency of size $x\in [a,b]$, and a strategy for the Invader is
to employ a width of frequency of size $y\in [a,c]$, where $c\leq
b<1/2$. Thus, we assume that the Invader's technical characteristics
(e.g., radio's capabilities) are not better than the Scanner's ones.

If the Scanner and the Invader use the strategies $x$ and $y$,
then the payoff to the Invader is the expected reward (which is a
function $U(y)$ of bandwidth $y$ illegally used by the Invader)
minus intrusion expenses (which is a function $C_I(y)$ of
bandwidth $y$) and expected fine $F$ to pay, i.e.,
\begin{equation}
\label{eE_a_0} v_I(x,y)= (1-P(x,y))U(y)-FP(x,y) -C_I(y).
\end{equation}

The Scanner wants to detect intrusion  taking into account
scanning the expenses and damage caused by the illegal use of the
bandwidth by the Invader. For detection, he is rewarded by a fine $F$ imposed on
the Invader. Thus, the payoff to the Scanner is the difference between
the expected reward for detection, and the damage from intrusion into
the bandwidth (which is a function $V(y)$ of bandwidth $y$
illegally used by the Invader) with  the scanning expenses (which
is a function $C_S(x)$ of scanned bandwidth $x$),
\begin{equation}
\label{eE_b} v_S(x,y)=FP(x,y)-V(y)(1-P(x,y))-C_S(x).
\end{equation}
Note that introducing transmission costs in such a formulation is common for CDMA
 \cite{ZSPB2011,AAG2010_WP} and ALOHA networks (\cite{SE2009,GHAA2012}).

\subsubsection{Incomplete information of
the Invader's reward and technical characteristics}

In this section we assume that the Invader's reward is defined by
the reason he intruded into the bandwidth illegally for, and
we consider two such reasons:

\begin{description}

\item[(a)] With probability $1-q_0$ for the Invader it is just
important to work in the network without being detected. Thus, if he
is not detected his reward is $U$ which does not depend on the width
of bandwidth employed for the intrusion.
 Then, of course, to minimize the probability of detection he will employ
 the minimal bandwidth allowed,
thus, his strategy is $y=a$.

\item[(b)] With probability $q_0$ for the Invader  the bandwidth he uses is important. Thus, his reward is
the same as in Section~\ref{sec:Invader reward is related to the
bandwidth used}. We assume that his technical characteristics can
be different, the Invader knows his characteristics, but the
Scanner does not know them. Under Invader's technical
characteristics we assume an  upper bound on the spectrum width he can employ.  The Scanner knows only this upper bound on bandwidth
 as $c$ with a conditional probability $q(c)\geq 0$ for
$c\in [a,b]$, i.e., $\int_{a}^b q(c)\,dc =1.$
\end{description}

To deal with this situation we are going to apply a Bayesian
approach, namely, we introduce  type $c\in [a,b]$  for the Invader
related to the corresponding upper bounds (thus, we here deal with
a continuum of Invader's types). The Invader knows his type, while the Scanner
knows only its distribution. Denote by $\by(c)\in [a,c]$
the strategy  of the Invader of type $c$. Then his payoff is given
as follows:
\begin{equation}
\label{eE_a_0_bay}
\begin{split}
 \bv^c_I(x,\by(c))&=
(1-P(x,\by(c)))U(\by(c))\\
&-FP(x,\by(c))-C_I(\by(c)).
\end{split}
\end{equation}
The payoff to the Scanner is the expected payoff taking into
account the \emph{type} of Invader:
\begin{equation}
\label{eE_b_0_bay}
\bv^E_S(x,\by)=(1-q_0)v_S(x,a)+q_0\int_{a}^bq(c) v_S(x,\by(c))\,dc
\end{equation}
with $v_S(x,\by(c))$ given by Eq. (\ref{eE_b}).

Here we look  for Bayesian equilibrium \cite{O1982}, i.e., for such couple of
strategies $(x_*,\by_*)$  that for any $(x,\by)$ the following
inequalities hold:
\begin{equation}
\label{e_BE}
\begin{split}
\bv^E_S(x,\by_*)&\leq \bv^E_S(x_*,\by_*),\\
\bv^c_I(x_*,\by(c))&\leq \bv^c_I(x_*,\by_*(c)),  c\in \mbox{supp}(q),
\end{split}
\end{equation}
with $\mbox{supp}(q)=\{c\in [a,b]: q(c)>0\}.$

We assume that the Scanner and the Invader know (as in the case
with complete information) the parameters $F$, $C_I$, $C_S$, $V$,
$U$, $a$, $b$ as well as the probabilities $q(c)$ ($c\in [a,b]$)
and $q_0$.

\section{\label{sec:Saddle point} Equilibrium  strategies for the first game}
In the following theorem we give the equilibrium  strategies for
the first game (thus, maxmin scanning algorithm) for fixed bound width of the rivals.

\begin{Theorem}
\label{th_1} In the first game with fixed  width to scan $x$ and
to invade $y$, the rivals employ a uniform tiling behavior. Namely,

(a) Let $1-(x+y)M\leq y$ with
\begin{equation}
\label{e_N} M=\left\lfloor 1/(x+y)\right\rfloor,
\end{equation}
where $\lfloor \xi \rfloor$ is the greatest integer less or equal
to $\xi$. Then the Scanner and the Invader will, with equal
probability $1/M$, employ a band of the set $A_{-S}$ and $A_{-I}$
correspondingly.

(b) Let $1-(x+y)M> y$. Then the Scanner and the Invader will, with
equal probability $1/(M+1)$, employ a band of the set $A_{+S}$ and
$A_{+I}$ correspondingly, where

\small
\begin{equation*}
\begin{split}
A_{-S}&=\{[k(x+y)-x,k(x+y)], k=1,...,M\},\\
A_{-I}&=\{[k(x+y)-y-\epsilon (M+1-k),k(x+y)-\epsilon (M-k)],\\
&k=1,..., M\}, \quad 0<\epsilon<x/M,
\\
A_{+S}&=A_{-S}\cup [1-x,1],
\\A_{+I}&=\{[(k-1)(x+y+\epsilon), (k-1)(x+y+\epsilon)+y],\\
&k=1,..., M\}\cup [1-y,1], \quad
0<\epsilon<\frac{1-y-M(x+y)}{M-1}.
\end{split}
\end{equation*}
\normalsize

\noindent The value of the game (detection probability) $P(x,y)$
is given as follows:
\begin{equation}
\label{eP_0} P(x,y)=
\begin{cases}
1/M,& 1-(x+y)M\leq y,\\
 1/(M+1),& 1-(x+y)M > y.\\
\end{cases}
\end{equation}

\end{Theorem}

\section{\label{sec:Second Step}Equilibrium strategy for the second game}
In this section, which is split into five subsections, we find the
equilibrium strategies for the second game explicitly. First, in
Subsection~\ref{sec:Linear} we linearize our model to get an
explicit solution, then in Subsection~\ref{sec:Best Response_2}
the best response strategies are given, and they are employed in
Subsections~~\ref{sec:Nash Equilibrium_Unknown} and \ref{sec:Nash
Equilibrium} to construct equilibrium strategies for known and
unknown Invader's technical characteristics correspondingly.

\subsection{\label{sec:Linear} Linearized  model}

In order to get an insight into the problem, we consider a
situation where the detection's probability $P(x,y)$ for $x,y\in
[a,b]$  is approximated by a linear function as follows:
\begin{equation}
\label{eP}
\begin{split}P(x,y)=x+y.
\end{split}
\end{equation}
Thus, Eq. (\ref{eP_0}) and Eq. (\ref{eP}) coincide for $x+y=1/n, n=2, 3,...$
We assume that the scanning and intrusion cost as well as the
Invader's and Scanner's utilities are linear in the bandwidth
involved, i.e., $C_S(x)=C_Sx$, $C_I(y)=C_Iy$, $U(y)=Uy$, $V(y)=Vy$
where $C_S, C_I, U, V>0$. Then the payoffs to the Invader and the
Scanner, if they use strategies $x\in [a,b]$ and $y\in [a,c]$
($\by(c)\in [a,c]$) respectively, become:

\begin{description}
\item[(i)] For the known Invader's reward:
\begin{equation*}
\begin{split}
v_I(x,y)&=U(1-x-y)y-F(x+y)-C_Iy,\\
v_S(x,y)&= F(x+y)-Vy(1-x-y)-C_Sx,
\end{split}
\end{equation*}

\item[(ii)] For the unknown Invader's reward and technical characteristics:
\end{description}

\begin{equation*}
\begin{split}
\bv^c_I(x,\by(c))&=U(1-x-\by(c))\by(c)\\
&-F(x+\by(c))-C_I\by(c),\mbox{ for }c\in \mbox{supp}(q),\\
\bv^E_S(x,\by)&= q_0\int_{a}^b \Bigl[F(x+\by(\xi))\\&\,\,\,\,\,\,\,\,\,\,\,\,\,\,\,\,\,\,\,\,\,-V\by(\xi)(1-x-\by(\xi))\Bigr]q(\xi)\,d\xi\\
&+(1-q_0)\Bigl(F(x+a)-Va(1-x-a)\Bigr)-C_Sx.
\end{split}
\end{equation*}

Note that linearized payoffs have found extensive usage for a wide
array of problems in wireless networks \cite{SE2009, AAG2009_d, KRZ1999, KALKZ1999, GT2013}. Of course, such an approach simplifies
the original problem  and only gives  an approximated solution.
Meanwhile, it can also be very useful: sometimes it allows one to
obtain a solution explicitly, and allows one to look inside of the
structure of the solution as well as the correlation between
parameters of the system.

\subsection{\label{sec:Best Response_2} The best response strategies}

In this section, we give the best response strategies for the Scanner
and the Invader when the Invader's reward and technical
characteristics are unknown, i.e., such strategies that
$\mbox{BR}^E_S(\by)=\arg\max_x v^E_S(x,\by)$ and
$\mbox{BR}^c_I(x)=\arg\max_{\by(c)} v^c_I(x,\by(c))$.

\begin{Theorem}
\label{th_BR_2} In the second step of the considered game with
unknown Invader's reward and technical characteristics the Scanner
and the Invader have the best response strategies
$\mbox{BR}^E_S(\by)$ and $\mbox{BR}^c_I(x)$ given as follows:

\begin{equation}
\label{e_th_2_BR_2}
 \mbox{BR}^E_S(\by)=
\begin{cases}
a,& \bar{\by}<R_{q_0},\\
\mbox{any from } [a,b],&  \bar{\by}=R_{q_0},\\
b,&  \bar{\by}>R_{q_0},
\end{cases}
\end{equation}

\begin{equation}
\label{e_th_2_BR_1}
\begin{split} \mbox{BR}^c_I(x)=
\begin{cases}
c,&  c\leq L(x),\\
L(x),& a<L(x)<c,\\
a,& L(x)\leq a
\end{cases}
\end{split}
\end{equation}

\noindent
with
\begin{equation}
\label{e_th_BR_2_L}
\begin{split}
L(x)&=\frac{T-x}{2}, \\
T&= \frac{U-F-C_I}{U}, \\
R&= \frac{C_S-F}{V},\\
R_{q_0}&=\frac{C_S-F-(1-q_0)Va}{q_0V}=\frac{R-a(1-q_0)}{q_0}
\end{split}
\end{equation}
and
$$
\bar{\by}=\int_{a}^b q(\xi)\by(\xi)\,d\xi.
$$
\end{Theorem}

\subsection{\label{sec:Nash
Equilibrium_Unknown}  Equilibrium strategies: the unknown Invader's reward and
technical characteristics}
 The equilibrium  for the game exists since the payoff to the Scanner is linear in
$x$ and the payoff to the Invader of type $c$ is concave in
$\by(c)$.    The equilibrium can be found by Eq. (\ref{e_BE}) as a
couple of strategies $(x,\by)$ which are the best response to each
other, i.e., $x=\mbox{BR}^E_S(\by)$ and $\by(c)=\mbox{BR}^c_I(x)$,
$c\in [a,b]$ and such a solution always exists and is unique as
shown in the following theorem.

\begin{Theorem}
\label{th_4} The considered second game with unknown Invader's reward and
technical characteristics has unique Bayesian equilibrium
$(x,\by)$, and it is given as follows:

\small
\begin{equation}
\label{e_th_4_xy}
\begin{split}
x&=
\begin{cases}
b,&R_{q_0}\leq \overline{\mbox{\bf BR}}_I(b),\\
\overline{\mbox{\bf
BR}}^{-1}_I\left(R_{q_0}\right),&\overline{\mbox{\bf
BR}}_I(b)<R_{q_0}<
\overline{\mbox{\bf BR}}_I(a),\\
a,& \overline{\mbox{\bf BR}}_I(a)\leq R_{q_0},
\end{cases}\\
\by(c)&=
\begin{cases}
\mbox{BR}^c_I(b),&R_{q_0}\leq \overline{\mbox{\bf BR}}_I(b),\\
\mbox{BR}_I^c\left(\mbox{\bf
BR}^{-1}_I(R_{q_0})\right),&\overline{\mbox{\bf BR}}_I(b)<R_{q_0}<
\overline{\mbox{\bf BR}}_I(a),\\
\mbox{BR}_I^c(a),& \overline{\mbox{\bf BR}}_I(a)\leq R_{q_0},
\end{cases}
\end{split}
\end{equation}
\normalsize \noindent where $c\in\mbox{supp}(q)$ with
\begin{equation}
\label{e_th_4_N}
\begin{split}
\overline{\mbox{\bf BR}}_I(x)=\int_{a}^b
q(\xi)\mbox{BR}^\xi_I(x)\,d\xi
\end{split}
\end{equation}
and $\overline{\mbox{\bf BR}}^{-1}_I(x)$ is inverse function to
$\overline{\mbox{\bf BR}}_I(x)$, i.e., $\overline{\mbox{\bf
BR}}^{-1}_I\left(\overline{\mbox{\bf BR}}_I(x)\right)=x.$
\end{Theorem}

\subsection{\label{sec:Nash Equilibrium}  Equilibrium strategies: the known Invader's technical
characteristics and  unknown reward}

The equilibrium  for the second game with complete
information about the technical characteristics of the Invader and
unknown reward can be presented explicitly  as follows:
\begin{Theorem}
\label{th_2} Let the Invader's technical characteristics be known
but his reward can be unknown.  This second game has unique Nash
equilibrium, and it is given by
Table~\ref{Tbl1}.

\begin{center}
\begin{table*}[ht]
{\small \hfill{}
\begin{tabular}{|c|c|c|c|c|c|c|}
\hline\hline
\textbf{Case}&\textbf{Condition}&\textbf{Condition}&$x$&$y$&
$P_R$&$P_U$\\
\hline\hline
$i_1$&$R_{q_0}<a$&$L(b)<a$&$b$&$a$&$a+b$&$2a$\\\hline
$i_2$&$R_{q_0}<a$&$a\leq L(b)\leq
c$&$b$&$L(b)$&$b+L(b)$&$b+a$\\\hline $i_3$&$R_{q_0}<a$&$c<
L(b)$&$b$&$c$&$b+c$&$b+a$\\\hline
$i_4$&$c<R_{q_0}$&$L(a)<a$&$a$&$a$&$2a$&$2a$\\\hline
$i_5$&$c<R_{q_0}$&$a\leq L(a)\leq
c$&$a$&$L(a)$&$a+L(a)$&$2a$\\\hline $i_6$&$c<R_{q_0}$&$c<
L(a)$&$a$&$c$&$a+c$&$2a$\\\hline $i_7$&$a\leq R_{q_0}\leq
c$&$L(b)\leq R_{q_0}\leq
L(a)$&$L^{-1}\left(R_{q_0}\right)$&$R_{q_0}$&$L^{-1}\left(R_{q_0}\right)+R_{q_0}$&$L^{-1}\left(R_{q_0}\right)+a$\\\hline
$i_8$&$a\leq R_{q_0}\leq c$&$L(a)\leq a$&$a$&$a$&$2a$&$2a$\\\hline
$i_9$&$a\leq R_{q_0}\leq
c$&$a<L(a)<R_{q_0}$&$a$&$L(a)$&$a+L(a)$&$2a$\\\hline
$i_{10}$&$a\leq R_{q_0}\leq
c$&$c<L(b)$&$b$&$c$&$b+c$&$b+a$\\\hline $i_{11}$&$a\leq
R_{q_0}\leq
c$&$R_{q_0}<L(b)<c$&$b$&$L(b)$&$b+L(b)$&$b+a$\\\hline\hline
\end{tabular}
} \hfill{}  \caption{\label{Tbl1} The equilibrium strategies
$(x,y)$ with $L^{-1}(R_{q_0})=T-2(C_S-F-(1-q_0)V)/(q_0V)$ and
$P_R$ and $P_U$ are detection probabilities of the Invader with
reward related and un-related to the bandwidth used.}
\end{table*}
\end{center}
\end{Theorem}

Note that the Scanner's and Invader's equilibrium strategies can
have sudden jumps (discontinuities) as one continuously varies the  fine $F$ and
probability $q_0$ that the Invader's reward related bandwidth
used. It is caused by the fact that $R_{q_0}$ depends on these
parameters, while $L$ depends only on $F$. For example,
($i_1$)-($i_6$) implies that the Invader's equilibrium strategy
can jump while probability $q_0$ varies, and ($i_2$) and ($i_6$)
yield about the possibility of such a jump by fine $F$. The possibility of
jumps for the Scanner's equilibrium strategy follows from  ($i_2$)
and ($i_5$).

\section{\label{sec:Numerical illustrations}Numerical illustrations}

\begin{figure}
\centering
\begin{tabular}{c}
\includegraphics[width=0.38\textwidth]{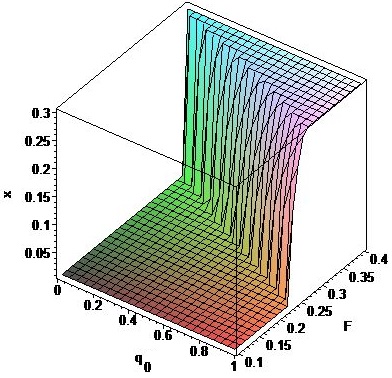}\\
\includegraphics[width=0.38\textwidth]{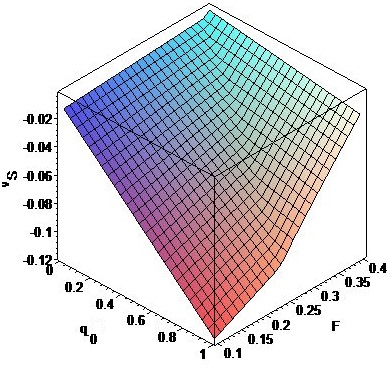}
\end{tabular}
\caption{\label{Fig_1}  Equilibrium strategy $x$ (upper) and payoff
(bottom) to the Scanner.}
\end{figure}

As a numerical illustration of the scenario with complete
information on the Invader's technical characteristics,  we consider
$U=V=1$, $a=0.01$, $b=0.3$,  $C_S=0.4$, $C_I=0.1$ and $q$ is the
uniform distribution in $[a_0,b_0]=[a+(b-a)/10,b]=[0.039,0.3]$.
Figure~\ref{Fig_1} demonstrates the Scanner's equilibrium
 strategy and payoff as functions of the fine $F\in
[0.1,0.4]$ and the probability $q_0\in [0.01,0.99]$ that the
Invader's reward related to bandwidth used. Increasing fine $F$
and probability  $q_0$ makes the Scanner employ a larger band and
impacts the Scanner's payoff in a multi-directional way, namely,
it increases $F$ and decreases $q_0$. This is caused by the
fact that the Invader, who wants to minimize his detection
probability, causes less damage to the network  than the one who
benefits from using a larger bandwidth.

\begin{figure}
\centering
\begin{tabular}{c}
\includegraphics[width=0.38\textwidth]{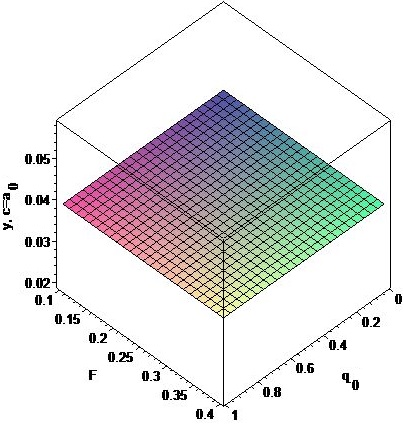}\\
\includegraphics[width=0.38\textwidth]{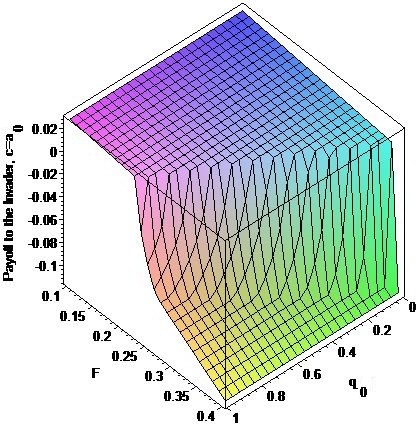}
\end{tabular}
\caption{\label{Fig_2}  Equilibrium strategy $y$ (upper) and payoff
(bottom) to the Invader for $c=a_0$.}
\end{figure}
\begin{figure}
\centering
\begin{tabular}{c}
\includegraphics[width=0.38\textwidth]{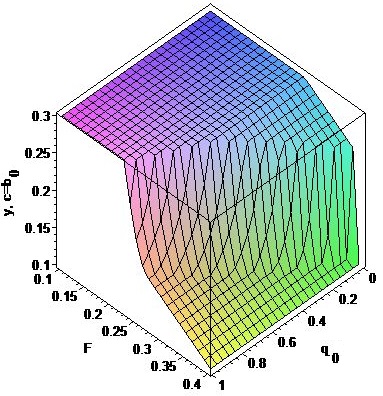} \\
\includegraphics[width=0.38\textwidth]{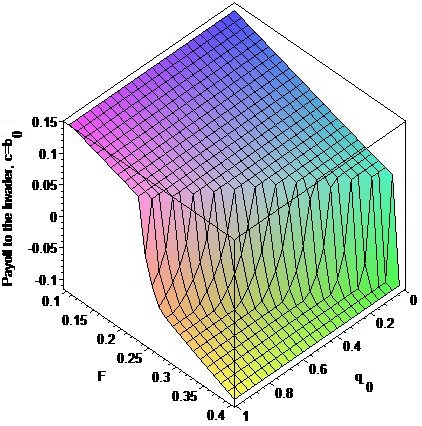}
\end{tabular}
\caption{\label{Fig_3}  Equilibrium strategy $y$ (upper) and payoff
(bottom) to the Invader for $c=b_0$.}
\end{figure}
\begin{figure}
\centering
\begin{tabular}{c}
\includegraphics[width=0.38\textwidth]{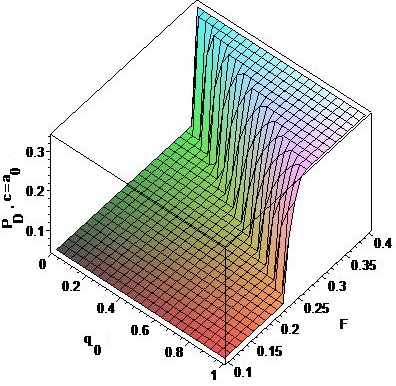}\\
\includegraphics[width=0.38\textwidth]{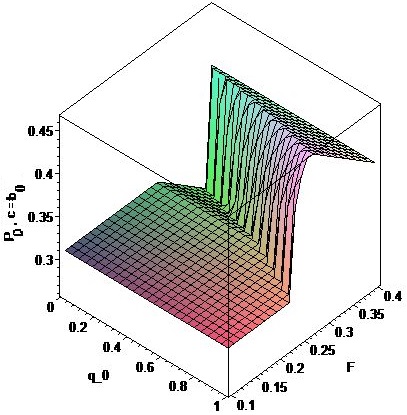}
\end{tabular}
\caption{\label{Fig_4}  Detection probability of the Invader with
$c=a_0$ (upper) and $c=b_0$ (bottom).}
\end{figure}
\begin{figure}
\centering
\begin{tabular}{cc}
\includegraphics[width=0.38\textwidth]{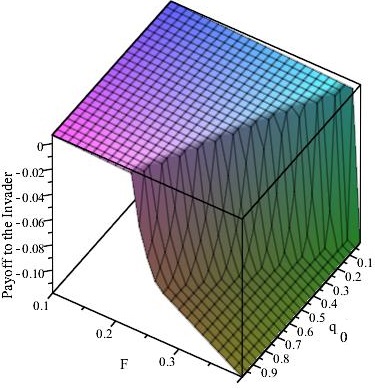}\\
\includegraphics[width=0.38\textwidth]{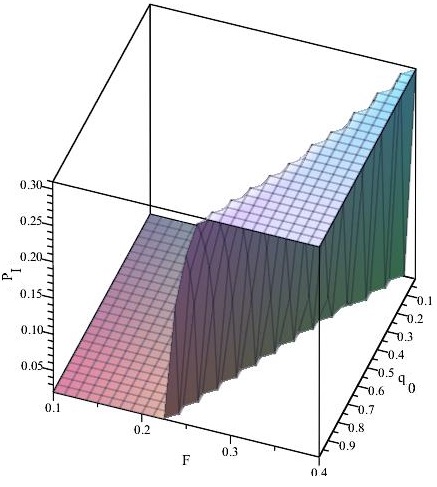}
\end{tabular}
\caption{\label{Fig_5}  Payoff (upper) and detection probability
(bottom) of the Invader with reward un-related  to the bandwidth
used.}
\end{figure}

Figures~\ref{Fig_2} and \ref{Fig_3} illustrate the Invader's
equilibrium strategy and payoff if his reward is related  to the
bandwidth used for $c=a_0$ and $c=b_0$ respectively. Figure~\ref{Fig_4}
demonstrates corresponding detection probabilities. The Invader of
type $c=a_0$ employs a constant strategy $y(c)=a_0$ independent of the
fine $F$ and probability $q_0$. The Invader's payoff and detection
probability vary in opposite directions while fine $F$ and
probability $q_0$ are increasing, namely, the Invader's payoff is
decreasing, while the detection probability is increasing, since
it also makes the Scanner to employ a larger bandwidth. What is
interesting is that the Invader's payoff experiences a sudden drop
and the detection probability experiences a sudden jump due to the
Scanner's behaviour, who alters his strategy by a sudden jump at
threshold values. For the Invader with a reward un-related to the
bandwidth used, the payoff and detection behave similarly but with some
shift since such an Invader also employs a constant strategy $y=a$
(Figure~\ref{Fig_5}). The Invader of type $c=b_0$ uses a strategy
depending on fine $F$ and probability $q_0$. Increasing fine $F$
and probability $q_0$ makes the Invader employ a smaller
bandwidth and reduces his payoff. What is interesting that his
detection probability is not monotonous by fine $F$ and
probability $q_0$ and increasing fine $F$ and probability $q_0$
could even reduce the detection probability. It can be explained
that at the threshold values of fine $F$ and probability $q_0$ the
Scanner already gets the upper band, while the Invader still does
not get to the lower band, and further increasing of the fine and
probability leads to continuous decreasing of the detection
probability due to the smaller bandwidth employed by the Invader.

\section{\label{sec:Discussion}Discussion}

In this paper, we suggest a simple model for designing
a maxmin scanning algorithm for detection of an Invader with incomplete
information about the Invader's reward and technical characteristics and we find the optimal parameters
(width of bandwidth to scan) for this algorithm.
We have shown that this optimal width essentially depends on the
network's and agent's characteristics and under some conditions a
small variation of network parameters and fine could lead to jump
changes in the optimal strategies, as well as in the payoffs of
the rivals. This mixture between continuous and discontinuous
behavior of the Invader under the influence of fine implies that
the network provider has to carefully make a value judgement: some
threshold values of fine could have a huge impact on the Invader,
while in the other situations a small increase will have a minimal
impact on the strategies used. A goal for our future
investigation is to investigate the non-linearized
detection probability. Also, we intend to extend our model to the
case of multi-step scanning algorithms with learning.





\section{Appendix}
\subsection{Proof of Theorem~\ref{th_1}}

Suppose that the Invader uses a band $B_I$ with width $y$ and the
Scanner with equal probability employ a band  from the set
$A_{-S}$ ($A_{+S}$) for $1-(x+y)M\leq y$ (for $1-(x+y)M>y$). The
intervals composing $A_{-S}$ and $A_{+S}$ are separated from each
other by at most $y$. So, at least one band from $A_{-S}$ for
$1-(x+y)M\leq y$ and from $A_{+S}$ for $1-(x+y)M> y$ intersects
with $B_I$.  So, detection probability is greater or equal to
$1/M$ for $1-(x+y)M\leq y$ and it is is greater or equal to
$1/(M+1)$ for $1-(x+y)M>y$.

Suppose that the Scanner uses a band $B_S$ with width $x$ and the
Invader with equal probability employ a band  from the set
$A_{-I}$ ($A_{+I}$) for $1-(x+y)M\leq y$ (for $1-(x+y)M>y$). The
intervals composing $A_{-I}$ and $A_{+I}$ are separated from each
other by more that $x$. So, at most one band from $A_{-I}$ for
$1-(x+y)M\leq y$ and from $A_{+I}$ for $1-(x+y)M> y$ intersects
with $B_S$.  So, detection probability is less or equal to $1/M$
for $1-(x+y)M\leq y$ and it is is less or equal to $1/(M+1)$ for
$1-(x+y)M>y$ and the result follows.
 \cqfd

\appendix
\section{Proof of Theorem~\ref{th_BR_2}} Note that
\begin{equation*}
\begin{split}
\bv^E_S(x,\by)&=\bigl(F-C_S+q_0V\bar{\by}+(1-q_0)Va\bigr)x\\
&+q_0\left[(F-V)\bar{\by}+V\int_{a}^b\by^2(\xi)q(\xi)\,d\xi\right]\\
&+(1-q_0)(F-V+Va)a.
\end{split}
\end{equation*}
So, for a fixed $\by$ the payoff $\bv^E_S(x,\by)$ is linear on
$x$. Thus, $\mbox{BR}^E_S(\by)=\arg\max_x \bv^E_S(x,\by)$ is
defined by sign of $F-C_S+q_0V\bar{\by}+(1-q_0)Va$ as it is given
by (\ref{e_th_2_BR_2}).

Note that, the Invader's payoff has the following form:
\begin{equation*}
\bv^c_I(x,\by(c))=(U(1-x)-F-C_I)\by(c)-U\by^2(c)-xF.
\end{equation*}
So, for a fixed $x$ the payoff $\bv^c_I(x,\by(c))$ is a concave
quadratic polynomial on $\by(c)$ getting its absolute maximum at
$\by(c)=(U(1-x)-F-C_I)/(2U)$. Thus, the maximum of
$\bv^c_I(x,\by(c))$ by $\by(c)$ within $[a,c]$ is reached either
on its bounds $\by(c)=a$ and $\by(c)=c$ or at
$\by(c)=(U(1-x)-F-C_I)/(2U)$ if it belongs to $[a,c]$ as it is
given by (\ref{e_th_2_BR_1}). \cqfd

\subsection{Proof of Theorem~\ref{th_4}}

First note that $(x,\by)$ is a Nash equilibrium if and only if it
is a solution of equations $x=\mbox{BR}^E_S(\by)$ and
$\by(c)=\mbox{BR}^c_I(x)$, $c\in [a,b]$ with $\mbox{BR}^E_S(\by)$
and $\mbox{BR}^c_I(x)$ given by Theorem~\ref{th_BR_2}.

By (\ref{e_th_BR_2_L}) we have that (\ref{e_th_2_BR_2}) in
equilibrium point  is equivalent to

\begin{equation}
\label{e_Equiv_2}
\begin{split} x=
\begin{cases}
a,& \overline{\mbox{\bf BR}}_I(x)< R_{q_0},\\
\mbox{any from } [a,b] ,&\overline{\mbox{\bf BR}}_I(x)=R_{q_0},\\
b,& \overline{\mbox{\bf BR}}_I(x)> R_{q_0}
\end{cases}
\end{split}
\end{equation}
with $\overline{\mbox{\bf BR}}_I(x)$ given by (\ref{e_th_4_N}).

Note that $\overline{\mbox{\bf BR}}_I(x)$ is non-increasing on
$x$. Thus,  if $\overline{\mbox{\bf BR}}_I(a)< R_{q_0}$, then
$\overline{\mbox{\bf BR}}_I(x)< R_{q_0}$ for any $x$ and
(\ref{e_Equiv_2}) yields that $x$ has to be equal to $a$. If
$\overline{\mbox{\bf BR}}_I(b)>R_{q_0}$, then $\overline{\mbox{\bf
BR}}_I(x)>R_{q_0}$ for any $x$ and (\ref{e_Equiv_2}) yields that
$x$ has to be equal to $b$. If $\overline{\mbox{\bf BR}}_I(b)\leq
R_{q_0} \leq \overline{\mbox{\bf BR}}_I(a)$ then
$x=\overline{\mbox{\bf BR}}^{-1}_I\left(R_{q_0}\right)$ and the
result follows.\cqfd

\subsection{Proof of Theorem~\ref{th_2}}
For the situation with complete information of the Invader's technical
characteristics the best response strategies turn into

\begin{equation}
\label{e_th_BR_1}
 \mbox{BR}_S(y)=
\begin{cases}
a,& y<R_{q_0},\\
\mbox{any from } [a,b],&  y=R_{q_0},\\
b,&  y>R_{q_0},
\end{cases}
\end{equation}

\begin{equation}
\label{e_Equiv}
\begin{split} \mbox{BR}_I(x)&=
\begin{cases}
a,& L(x)\leq a,\\
L(x),&a<L(x)<c,\\
c,&  c\leq L(x)
\end{cases}\\
&=
\begin{cases}
a,& x\leq T-2c,\\
L(x),&T-2c<x<T-2a,\\
c,&  T-2a\leq x.
\end{cases}
\end{split}
\end{equation}

\begin{figure}
\centering
\includegraphics[width=0.24\textwidth]{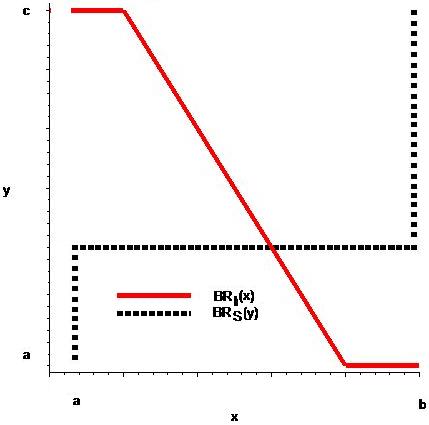}
\caption{\label{Fig_0} The Nash equilibrium as an intersection of
the best response curves}
\end{figure}

Thus, the equilibrium can be described as an intersection of the
best response curves (Figure~\ref{Fig_0}). Such intersection
always exists.

Let $a>R_{q_0}$. By (\ref{e_th_BR_1}), $\mbox{BR}_S(y)\equiv b$.
This, jointly with (\ref{e_Equiv}),  implies ($i_1$)-($i_3$).

Let $R_{q_0}>c$. By (\ref{e_th_BR_1}), $\mbox{BR}_S(y)\equiv a$.
Then, (\ref{e_Equiv})  implies ($i_4$)-($i_6$).

Let $a\leq R_{q_0}\leq c$. First note $L(x)$ is linear
decreasing function from $L(a)$ for $x=a$ to $L(b)$ for $x=b$.

\begin{description}
\item[(a)] Let $L(b)\leq R_{q_0}\leq L(a)$. Then the equation
$L(x)=R_{q_0}$ has the unique root within $[a,b]$. Thus,
(\ref{e_th_BR_1}) and (\ref{e_Equiv}) yield ($i_7$).

\item[(b)] Let $L(a)\leq R_{q_0}$. Then, $L(x)<R_{q_0}$ for $x\in (a,b]$.
So, by (\ref{e_Equiv}), $\mbox{BR}_I(x)<c$ for $x\in [a,b]$.
Besides, by the assumption, the equation $L(x)=R_{q_0}$ does not
has root in $[a,b]$. Thus, by (\ref{e_th_BR_1}),
$\mbox{BR}_S(y)\equiv a$. So, (\ref{e_Equiv})  implies ($i_8$) and
($i_9$).

\item[(c)] Let $R_{q_{0}}<L(b)$. Then $L(x)>R_{q_0}$ for $x\in [a,b)$.
Thus, by (\ref{e_Equiv}), $\mbox{BR}_I(x)>a$ for $x\in [a,b]$.
Besides, by the assumption, the equation $L(x)=R_{q_0}$ does not
has root in $[a,b]$. So, by (\ref{e_th_BR_1}), $\mbox{BR}_S(y)=b$,
and, (\ref{e_Equiv})  implies ($i_{10}$) and ($i_{11}$).
\end{description}
\cqfd

\bibliographystyle{ieeetr}
\bibliography{bib_Crowncom_2013}

\end{document}